\RequirePackage{fix-cm}
\documentclass[twocolumn]{svjour3}

\smartqed

\usepackage{graphicx}
\usepackage{amssymb}
\usepackage{caption}
\usepackage{subcaption}
\usepackage{hyperref}
\usepackage{todonotes}
\usepackage{booktabs}
\usepackage{multirow}
\usepackage[binary-units=true]{siunitx}
\DeclareSIUnit\flop{F}

\usetikzlibrary{arrows.meta, positioning, quotes, calc, intersections, positioning, matrix, shapes.geometric, shapes.multipart}

\newsavebox{\twosubbox}

\usepackage[sorting=none,style=numeric-comp]{biblatex}
\journalname{Computing and Software for Big Science}

\begin{document}
\emergencystretch 3em
\title{Fast Columnar Physics Analyses of Terabyte-Scale LHC Data on a Cache-Aware Dask Cluster}

\author{Niclas Eich \and Martin Erdmann \and Peter Fackeldey \and Benjamin Fischer \and Dennis Noll \and Yannik Rath}

\institute{\at
              Physics Institute 3A, RWTH Aachen University, 52056 Aachen, Germany \\
              Tel.: +49-241-80-27330\\
              Fax:  +49-241-80-22189\\
              \email{peter.fackeldey@rwth-aachen.de}
}

\date{Received: date / Accepted: date}

\maketitle

\begin{abstract}
The development of an LHC physics analysis involves numerous investigations that require the repeated processing of terabytes of data. Thus, a rapid completion of each of these analysis cycles is central to mastering the science project.
We present a solution to efficiently handle and accelerate physics analyses on small-size institute clusters.
Our solution is based on three key concepts: Vectorized processing of collision events, the ``MapReduce'' paradigm for scaling out on computing clusters, and efficiently utilized SSD caching to reduce latencies in IO operations.
Using simulations from a Higgs pair production physics analysis as an example, we achieve an improvement factor of $6.3$ in runtime after one cycle and even an overall speedup of a factor of $14.9$ after $10$ cycles.
\keywords{Data analysis \and Scaling \and Cache \and High-throughput computing}
\end{abstract}

\section{Introduction}
\label{intro}
Obtaining new physics results from LHC collider data at CERN ranks among the most challenging tasks in data analyses. Although the raw data of the experiments are processed centrally and quantities such as particle tracks are reconstructed, numerous tasks remain for small analysis teams to achieve a concrete scientific result. An example is the cross section measurement of the Higgs boson production.

When developing a physics analysis, many different studies need to be performed. Examples include data-driven background estimations, efficiency measurements, training and evaluation of multivariate methods, and determination of systematic uncertainties. All of these studies typically require multiple processing of at least a significant portion of the data and simulations.
In addition, analyses are subjected to an experiment-internal peer review process, which requires numerous further consolidation studies. Consequently, every data analysis is inevitably subjected to a large number of iterations.

Two challenges are of central importance: First, to perform the analysis in a reproducible manner, we rely on a workflow management system for data analysis~\cite{rieger2017design}. Second, the turnaround time of each analysis cycle is critical to making progress. In the best case, the runtime of the analysis cycle harmonizes with the reflection phase, in which the physicist decides on the next action. 

The duration of an analysis cycle has considerably increased due to the very successful LHC operation and the associated growth of recorded data. Typical analyzed data volumes are in the order of terabytes (TB). Without further developments in analysis technology, the prospect of the LHC upgrade for high luminosities will again significantly prolong analysis cycles. Three key concepts are exploited in this work to compensate this increase and improve the runtime of an analysis cycle.

The first concept tackles the way of processing events. While classically collision events are analyzed one after another, vectorized array operations can process multiple events simultaneously. The scientific Python ecosystem NumPy~\cite{harris2020array} provides these vectorized array operations using the processor-specific `single instruction multiple data' (SIMD) instruction sets.

Secondly, the programming paradigm `MapReduce'~\cite{mapreduce} is a key concept for this project. Operations, such as selection and reconstruction, are mapped to subsets of collision events. Their partial output is then accumulated (reduced) to a single output. Software packages such as Dask~\cite{dask} orchestrate this paradigm on any computing cluster.

The third key concept is caching. Caching increases the efficiency of repeated data access. Here, we present a caching mechanism that caches collision data on processor-near solid-state disks (SSDs). Subsequent analysis cycles benefit from this and show a substantial reduction in cycle time.

For the first two key concepts, there are already established software solutions, e.g., NumPy and Dask, that aim at accelerating computationally intensive operations by means of parallelization. With this speedup we uncovered a new limitation that is addressed by the third key concept: The time spent on IO operations, especially transferring the collision data to the processors, accounts for a non-negligible portion of the total runtime.

This paper presents a benchmark using modern analysis technologies for speeding up analysis cycles through the previously introduced key concepts. For the first two concepts, we employ the coffea~\cite{coffea} and Dask software packages. As resources, we use the computing cluster of the VISPA project, which provides cloud services for scientific data analysis~(\cite{Erdmann:2019aiy} and references therein). For the third concept, we have substantially extended the computing cluster with solid-state storage disks. As an application example, we present data reading benchmarks using simulated collision events of a Higgs pair production analysis.

This work is structured as follows. We describe the upgraded VISPA platform, quote the software components, explain the caching, and conduct a quantitative survey on the runtime reduction for multiple consecutive analysis cycles.

\section{VISPA Hardware and Software Systems}

\subsection{Cluster Setup}
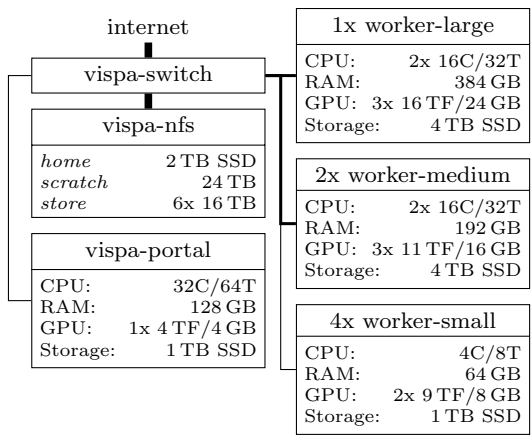
\begin{figure}[h]
    \centering
    \begin{tikzpicture}[
    node distance=2mm and 4mm,
    box/.style = {draw, text height=1.5ex,text width=9em, align=center},
    square/.style={draw, regular polygon,regular polygon sides=4}
    sy+/.style = {yshift= 2mm}, 
    sy-/.style = {yshift=-2mm},
    every edge quotes/.style = {align=center},
    pics/worker/.style n args={8}{code={%
        \node[draw,
        rectangle split,
        rectangle split parts=2,
        text height=1.5ex,
        text width=9em,
        text centered
        ] (#1)
        {
        #2 #1\nodepart[font=\scriptsize]{second}
        CPU: \hfill #3\\
        RAM: \hfill \SI{#4}{\giga\byte}\\
        GPU: \hfill #6 \SI{#7}{\tera\flop}/\SI{#8}{\giga\byte}
        Storage: \hfill \SI{#5}{\tera\byte} SSD\\
        };
    }},
    pics/servicemachine/.style n args={2}{code={%
        \node[draw,
        rectangle split,
        rectangle split parts=2,
        text height=1.5ex,
        text width=9em,
        text centered
        ] (#1)
        {
        #1\nodepart[font=\scriptsize]{second}%
        #2%
        };%
    }},
]
    \newcommand\HEIGHT{3.95cm}
    \newcommand{\edgelinerightone}[2]{
        \path(#1.east) --(#2.west)  coordinate[pos=0.5](mid);
        \draw (#1.east) -| (mid) |- (#2.west);
    }
    \newcommand{\edgelinerightfour}[2]{
        \path(#1.east) --(#2.west)  coordinate[pos=0.5](mid);
        \draw [line width=0.40mm] (#1.east) -| (mid) |- (#2.west);
    }
    \newcommand{\edgelineleft}[2]{
        \draw (#1.west) -- ($(#1.west)-(0.3, 0)$) -- ($(#2.west)-(0.3, 0)$) -- (#2.west);
    }

    \node (world) {internet};
    \node (switch) [box,below=of world] {vispa-switch};
    
    \pic [below=of switch] {servicemachine={vispa-nfs}{\emph{home} \hfill \SI{2}{\tera\byte} SSD\\ \emph{scratch} \hfill \SI{24}{\tera\byte} \\ \emph{store} \hfill 6x \SI{16}{\tera\byte}}};
    \pic [below=of vispa-nfs] {worker={vispa-portal}{}{32C/64T}{128}{1}{1x}{4}{4}}; % /SP3
    \pic [right=of switch] {worker={worker-large}{1x}{2x 16C/32T}{384}{4}{3x}{16}{24}}; % /FCLGA3647, Quadro RTX6000
    \pic [below=of worker-large] {worker={worker-medium}{2x}{2x 16C/32T}{192}{4}{3x}{11}{16}}; % /FCLGA3647, Quadro RTX5000
    \pic [below=of worker-medium] {worker={worker-small}{4x}{4C/8T}{64}{1}{2x}{9}{8}}; % /LGA1155, GTX1080
    
    \draw [line width=1mm] (world) -- (switch);
    \draw [line width=1mm] (switch) -- (vispa-nfs);
    \edgelineleft{switch}{vispa-portal};

    \edgelinerightfour{switch}{worker-large};
    \edgelinerightfour{switch}{worker-medium};
    \edgelinerightone{switch}{worker-small};

\end{tikzpicture}
    \caption{The hardware setup of the used cluster. A total of 10 different nodes are used of which three nodes are for service and 7 nodes are utilized as worker nodes. Central processing units (CPU) are specified with their number of cores (C) and the number of threads (T). The capabilities of the graphic processing units (GPU) are expressed in their floating-point performance (FP32) and memory (VRAM). Network connections are drawn by lines, whereas their width corresponds to the provided bandwidth (1/4/\SI[per-mode=symbol]{10}{\giga\bit\per\second}).}
    \label{fig:hardware_setup}
\end{figure}

The setup used in the presented analysis is a small-scale computing cluster that is optimized for scientific data analysis and deep learning applications (Fig.~\ref{fig:hardware_setup}).

It features various service nodes as well as three different sizes of worker nodes, which differ mostly in the processors (CPUs), the RAM capabilities, the graphics processing units (GPU), and their network connections.
The service node vispa-portal is used for interactive working and the management of the batch system (see section \ref{subsec:scale}).
It possesses a CPU with 64 logical cores and \SI{128}{\giga\byte} RAM.
The seven worker nodes have a combined CPU capacity of 224 logical cores and possess a total of \SI{832}{\giga\byte} RAM and \SI{16}{\tera\byte} SSD storage for caching purposes.
The detailed configurations of the individual machines can be found in Fig.~\ref{fig:hardware_setup}.
All used processors are capable of the AVX2 SIMD instruction set~\cite{avx2}.
The service node vispa-portal and each worker node additionally possess a \SI{4}{\tera\byte} SSD used for local storage and caching.
The caching strategy and its implementation is explained in section~\ref{subsec:cache}.

The switch (vispa-switch) is the central node of the local network.
It is connected to the internet and the storage service node (vispa-nfs) with \SI[per-mode=symbol]{10}{\giga\bit\per\second}, to the large and medium workers with \SI[per-mode=symbol]{4}{\giga\bit\per\second}, and to each small worker node and the service node vispa-portal with \SI[per-mode=symbol]{1}{\giga\bit\per\second}.
Additionally, each node has a fully isolated out-of-band management interface.

The service node vispa-nfs provides central storage capacity for the cluster.
It possesses a total of \SI{2}{\tera\byte} SSD and \SI{120}{\tera\byte} HDD storage.
The storage can be accessed via three different network-shared file systems implemented with the Network File System (NFS) protocol (version 4.2)~\cite{rfc7862}.
The file system \emph{home} is used as the working directories of the users.
It is mirrored and backed up daily by the local computing authority.
Additionally, two different file systems, \emph{scratch} and \emph{store}, can be used for larger amounts of data.
The \emph{scratch} file system, which totals to \SI{24}{\tera\byte}, is mirrored and used for experimental data and intermediate results, such as pre-processed experimental data.
The \emph{store} file system, which totals \SI{96}{\tera\byte}, has the purpose to store reproducible data, such as local copies of raw experimental data or software installations.
Because it is striped across 6 different \SI{16}{\tera\byte} HDDs it features fast reading and writing.
It is therefore suited for data, which is accessed frequently or with a high total throughput.
The total shared file system bandwidth is limited by the vispa-nfs network bandwidth of up to \SI[per-mode=symbol]{10}{\giga\bit\per\second}.

Operating systems are deployed on the different nodes using the open-source configuration management tool Ansible~\cite{ansible}.
It provides a declarative way to describe the configuration of the whole setup, ensuring reproducibility and thus supporting cluster stability.

The provisioning of the user's working environment is done using the open-source package management system conda~\cite{anaconda}. It ensures the stability and maintainability of each user's working environment, even in a heterogeneous and changing computing setup, and adapts to the multiple different needs of a large user base.

{}

\subsection{Job Distribution with HTCondor}
\label{subsec:scale}

Scaling analyses to run on the entire cluster requires a solution for workload management. While small jobs can be run interactively on the vispa-portal node, larger workflows are distributed to the worker nodes using HTCondor~\cite{htcondor}.

The HTCondor setup in VISPA consists of three main parts: a scheduler, a central manager, and workers.
Users submit their jobs to the HTCondor scheduler. Each of these jobs defines requirements that specify the resources it is expected to consume. The central manager then performs the matchmaking between these requirements and the available resources of the workers.

For the analysis presented here, the workload is split into chunks that can be distributed over the cluster using Dask and Dask-Jobqueue~\cite{dask-jq1,dask-jq2}.
The user launches a Dask scheduler on vispa-portal, and Dask workers are spawned on the worker nodes via HTCondor jobs. The Dask scheduler then distributes chunks of the total workload to these workers.
This distribution requires unrestricted communication among the Dask scheduler and the VISPA worker nodes.

\subsection{SSD Caching}
\label{subsec:cache}

Modern high-energy physics analyses need to analyze data on the terabyte scale. Using vispa-nfs for reading these data from \emph{scratch} is strongly limited by the HDDs and network connections.
This limitation is alleviated using appropriate caching mechanisms, as described in the following.

The caching is facilitated for each worker by the FSCache available within the Linux kernel~\cite{fscache}.
Once enabled for a particular NFS mount-point, it operates transparently upon all I/O requests for files therein.
In particular, data is cached at a page-size granularity (\SI{4}{\kilo\byte}) which enables selective caching, i.e., of only the accessed branches of a \texttt{.root} file.
Since all I/O operations (read and write) fill the cache, the occurrence of cache-trashing is minimized by only caching the \emph{store} volume, which is predominantly used for write-once read-often data.
The cache is configured to store its contents on the SSDs of the workers, thus profiting from their superior data transfer rates.

Since each cache will only contain the contents of data requested by its worker at some point prior, it is of utmost importance to route such requests - or rather the jobs that cause these particular requests - in a cache-hit maximizing manner.
This is done through a worker~--~job affinity mechanism, where each worker and job is identified in a reproducible manner. The identifier consists of
1) the worker by its hostname, 2) the job by the input file UUID, and 3) the range of the event numbers.

These identifiers are then uniformly mapped into a high dimensional bounded space by interpreting their cryptographic hash value (i.e., SHA512) as a vector of integers (i.e., $[0..255]^{64}$).
For any pair of such values a distance $D$ can be calculated as such: $D(\vec{a},\vec{b}) = \sum_i d(\left|a_i - b_i\right|)$ where $d(x) = \min(x,256-x)$.
Each job is then assigned to the worker it has the smallest distance to, ensuring a reasonably even distribution. The assignment is not strict, allowing idle workers to steal jobs from busy workers. This concept is referred to as work-stealing. It avoids trailing jobs due to heterogeneous job runtimes, thus improving the overall runtime.
Especially in the case of the addition or removal of workers, the affected jobs are redistributed homogeneously while avoiding the reallocation of all other jobs.
Additionally, the allocation ratio of jobs between workers can be changed smoothly by including a worker-specific distance factor - which is used to equalize the workload despite the varying processing power of all the workers.
Multiple users can participate and profit from the data caching when using the same files and affinity mechanism.

\section{Performance Benchmark}
\label{sec:benchmark}

\begin{figure}[h]
  \includegraphics[width=\linewidth]{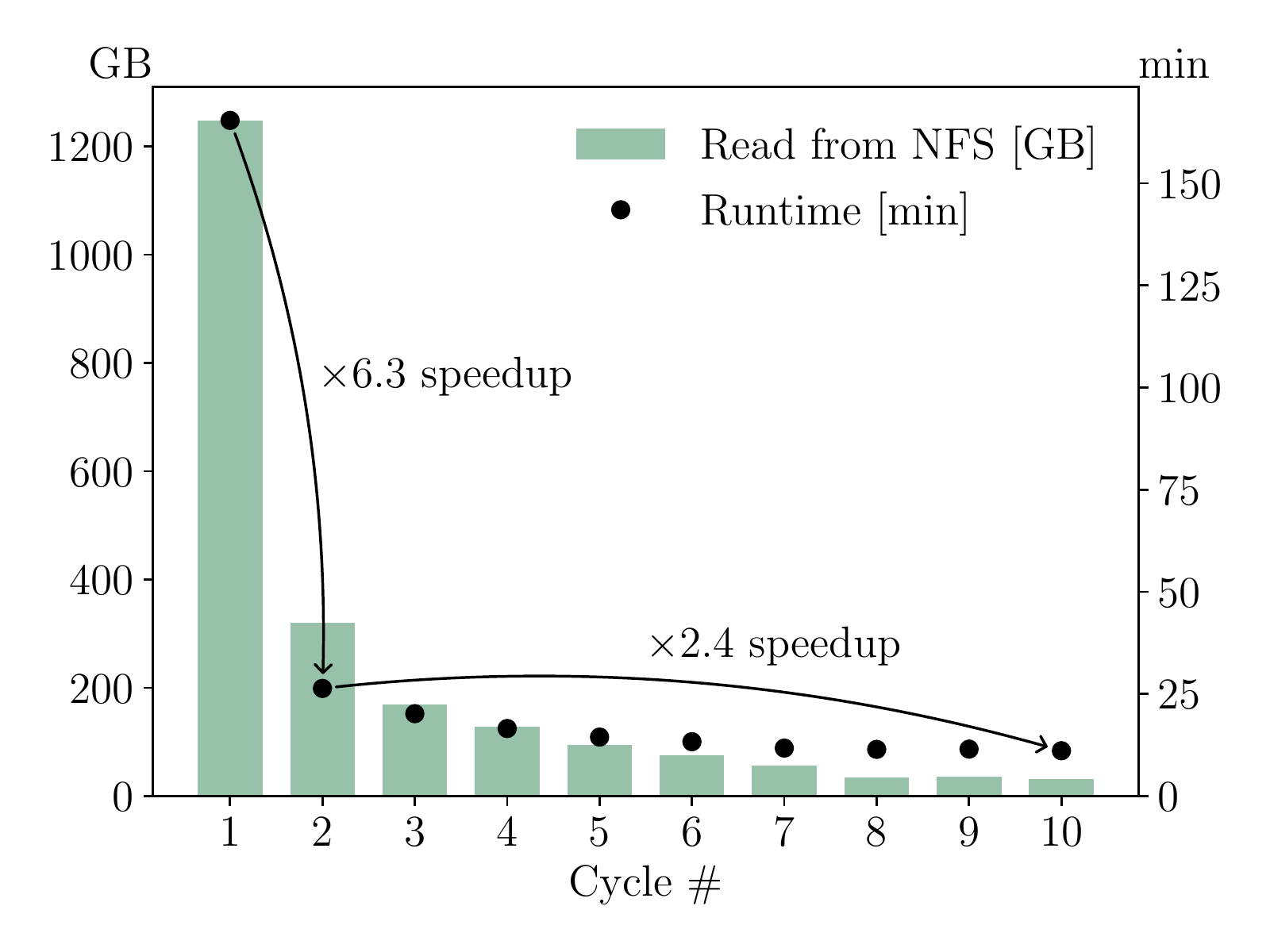}
\caption{Performance benchmark results for ten consecutive cycles.}
\label{fig:benchmark}
\end{figure}

The performance of the VISPA computing cluster with on-worker SSD caching is measured for a subset of simulated datasets in the NanoAOD data format~\cite{NanoAOD}. In total, the read data amounts to \SI{1439}{\giga\byte}, which corresponds to the event information of $\SI{1.05e9}{}$ events in the scope of a realistic Higgs pair production analysis. All datasets are compressed with the level ten Z-standard compression algorithm~\cite{zstd}, which has been changed from NanoAOD's default compression in order to reduce the decompression time. Our benchmark consists of multiple consecutive cycles. Throughout each cycle, 221 Dask workers carry out the processing with one thread and \SI{1.5}{\giga\byte} RAM each. Fig.~\ref{fig:benchmark} shows the performance benchmark for ten cycles.

The key message is that the runtime decreases substantially for the first few cycles. In total the improvement amounts to a factor of $14.9$. The amount of data that is still read from the vispa-nfs is vastly reduced as more data is read from the on-worker SSD caches. This effect converges for later cycles until almost all data is cached directly in the on-worker SSDs. The close overlay of runtime and the amount of data, which is still read from the vispa-nfs, show a strong correlation between runtime reduction and caching. The cache usage gradually converges to a maximum since a work-stealing mechanism minimizes each cycle's runtime at the cost of slight degraded deterministic cache usage.

\section{Conclusion}

Modern LHC physics analyses need to deal with a large amount of recorded data, while analysts and collaborations require many analysis cycles for various studies in the shortest time possible. Even physics analyses using vectorized processing of events and the MapReduce paradigm can significantly benefit from a dedicated on-worker SSD caching strategy. On the VISPA system, a small scale computing cluster, we show that a consequent caching strategy significantly reduces the time spent in IO operations. In the scope of a real-world Higgs pair production analysis, our benchmark shows a speedup of a factor of $14.9$ for the $10$th analysis cycle, moving from a few hours to approximately ten minutes. This allows numerous more analysis cycles and diminishes the IO limitation of large-scale analysis projects.

The caching strategy described in this paper allows for overcoming IO bottlenecks of modern LHC physics analyses, enabling small-scale computing clusters to become a competitive choice for interactive, flexible, and high-performance physics analyses.

\section*{Acknowledgments}

This work is supported by the Ministry of Innovation, Science, and Research of the State of North Rhine-Westphalia, and by the Federal Ministry of Education and Research (BMBF) in Germany. N.E. gratefully acknowledges the support of the Deutsche Forschungsgemeinschaft.

\section*{Literature}
\printbibliography[heading=none,notkeyword=hot]

@software{coffea,
  author       = {Lindsey Gray and
                  Nicholas Smith and
                  Benjamin Tovar and
                  Andrzej Novak and
                  Peter Fackeldey and
                  Nikolai Hartmann and
                  Jayjeet Chakraborty and
                  Gordon Watts and
                  Douglas Thain and
                  Giordon Stark and
                  BenGalewsky and
                  Devin Taylor and
                  bfis and
                  Jonas Rübenach and
                  Cami Carballo and
                  Dmitry Kondratyev and
                  Paul Gessinger and
                  Yi-Mu "Enoch" Chen and
                  Joosep Pata and
                  Anna Woodard and
                  Andreas Albert and
                  Zoe Surma and
                  Alexx Perloff and
                  Kevin Pedro and
                  dnoonan08 and
                  Karol Krizka and
                  kmohrman and
                  Lukas and
                  Massimiliano Galli and
                  Nick Amin},
  title        = {CoffeaTeam/coffea: Release v0.7.11},
  month        = dec,
  year         = 2021,
  publisher    = {Zenodo},
  version      = {v0.7.11},
  doi          = {10.5281/zenodo.5762406},
  url          = {https://doi.org/10.5281/zenodo.5762406}
}

@misc{rieger2017design,
      title={Design and Execution of make-like, distributed Analyses based on Spotify's Pipelining Package Luigi}, 
      author={Marcel Rieger and Martin Erdmann and Benjamin Fischer and Robert Fischer},
      year={2017},
      eprint={1706.00955},
      archivePrefix={arXiv},
      primaryClass={physics.data-an}
}

@article{NanoAOD,
	author = {{Ehat\"aht, Karl}},
	title = {NANOAOD: a new compact event data format in CMS},
	DOI= "10.1051/epjconf/202024506002",
	url= "https://doi.org/10.1051/epjconf/202024506002",
	journal = {EPJ Web Conf.},
	year = 2020,
	volume = 245,
	pages = "06002",
}

@Manual{dask,
  title = {Dask: Library for dynamic task scheduling},
  author = {{Dask Development Team}},
  year = {2016},
  url = {https://dask.org},
  urldate = {2022-05-26} 
}

@Misc{dask-jq1,
title = {dask-jobqueue source code},
urldate = {2022-05-26},
url = {https://github.com/dask/dask-jobqueue},
howpublished = "\url{https://github.com/dask/dask-jobqueue}",
  urldate = {2022-05-26} 
}

@Misc{dask-jq2,
title = {dask-jobqueue blog entry},
note = {accessed on 01.02.2022},
url = {https://blog.dask.org/2018/10/08/Dask-Jobqueue},
howpublished = "\url{https://blog.dask.org/2018/10/08/Dask-Jobqueue}",
urldate = {2022-05-26} 
}

@software{htcondor,
  author       = {HTCondor Team},
  title        = {HTCondor},
  month        = dec,
  year         = 2021,
  publisher    = {Zenodo},
  version      = {9.4.0},
  doi          = {10.5281/zenodo.5750673},
  url          = {https://doi.org/10.5281/zenodo.5750673}
}

@misc{anaconda, 
  title={Anaconda Software Distribution}, 
  url={https://docs.anaconda.com/}, 
  journal={Anaconda Documentation}, 
  version={Vers. 2-2.4.0},  
  publisher={Anaconda Inc.}, 
  year={2020}, 
  urldate = {2022-05-26} 
}

@misc{zstd,
	series =	{Request for Comments},
	number =	8878,
	howpublished =	{RFC 8878},
	publisher =	{RFC Editor},
	doi =		{10.17487/RFC8878},
	url =		{https://www.rfc-editor.org/info/rfc8878},
        author =	{Yann Collet and Murray Kucherawy},
	title =		{{Zstandard Compression and the 'application/zstd' Media Type}},
	pagetotal =	45,
	year =		2021,
	month =		feb,
}

@inproceedings{mapreduce,
title	= {MapReduce: Simplified Data Processing on Large Clusters},
author	= {Jeffrey Dean and Sanjay Ghemawat},
year	= {2004},
booktitle	= {OSDI'04: Sixth Symposium on Operating System Design and Implementation},
pages	= {137--150},
address	= {San Francisco, CA}
}

@article{Erdmann:2019aiy,
    author = "Erdmann, Martin and others",
    editor = "Forti, A. and Betev, L. and Litmaath, M. and Smirnova, O. and Hristov, P.",
    title = "{Evolution of the VISPA-project}",
    doi = "10.1051/epjconf/201921405021",
    journal = "EPJ Web Conf.",
    volume = "214",
    pages = "05021",
    year = "2019"
}

@online{ansible,
  author = {Red Hat},
  title = {Ansible},
  year = 2022,
  url = {https://www.ansible.com},
  urldate = {2022-05-26}
}

@online{fscache,
  author = {Linux Kernel},
  title = {General Filesystem Caching},
  year = 2022,
  url = {https://www.kernel.org/doc/html/latest/filesystems/caching/fscache.html},
  urldate = {2022-05-26}
}

@online{avx2,
  author = {Intel},
  title = {Advanced Vector Extensions},
  year = 2022,
  url = {https://www.intel.de/content/www/de/de/architecture-and-technology/avx-512-overview.html},
  urldate = {2022-05-26}
}

@misc{rfc7862,
	series =	{Request for Comments},
	number =	7862,
	howpublished =	{RFC 7862},
	publisher =	{RFC Editor},
	doi =		{10.17487/RFC7862},
	url =		{https://www.rfc-editor.org/info/rfc7862},
        author =	{Thomas Haynes},
	title =		{{Network File System (NFS) Version 4 Minor Version 2 Protocol}},
	pagetotal =	104,
	year =		2016,
	month =		nov,
}

@Article{         harris2020array,
 title         = {Array programming with {NumPy}},
 author        = {Charles R. Harris and K. Jarrod Millman and St{\'{e}}fan J.
                 van der Walt and Ralf Gommers and Pauli Virtanen and David
                 Cournapeau and Eric Wieser and Julian Taylor and Sebastian
                 Berg and Nathaniel J. Smith and Robert Kern and Matti Picus
                 and Stephan Hoyer and Marten H. van Kerkwijk and Matthew
                 Brett and Allan Haldane and Jaime Fern{\'{a}}ndez del
                 R{\'{i}}o and Mark Wiebe and Pearu Peterson and Pierre
                 G{\'{e}}rard-Marchant and Kevin Sheppard and Tyler Reddy and
                 Warren Weckesser and Hameer Abbasi and Christoph Gohlke and
                 Travis E. Oliphant},
 year          = {2020},
 month         = sep,
 journal       = {Nature},
 volume        = {585},
 number        = {7825},
 pages         = {357--362},
 doi           = {10.1038/s41586-020-2649-2},
 publisher     = {Springer Science and Business Media {LLC}},
 url           = {https://doi.org/10.1038/s41586-020-2649-2}
}

\end{document}